# FARe: Fault-Aware GNN Training on ReRAM-based PIM Accelerators

Pratyush Dhingra[1], Chukwufumnanya Ogbogu[1], Biresh Kumar Joardar[2], Janardhan Rao Doppa[1], Ananth Kalyanaraman[1], Partha Pratim Pande[1].
[1] Washington State University, Pullman WA, USA. [2] University of Houston, Houston TX, USA.

*Abstract*— Resistive random-access memory (ReRAM)-based processing-in-memory (PIM) architecture is an attractive solution for training Graph Neural Networks (GNNs) on edge platforms. However, the immature fabrication process and limited write endurance of ReRAMs make them prone to hardware faults, thereby limiting their widespread adoption for GNN training. Further, the existing fault-tolerant solutions prove inadequate for effectively training GNNs in the presence of faults. In this paper, we propose a fault-aware framework referred to as FARe that mitigates the effect of faults during GNN training. FARe outperforms existing approaches in terms of both accuracy and timing overhead. Experimental results demonstrate that FARe framework can restore GNN test accuracy by 47.6% on faulty ReRAM hardware with a ~1% timing overhead compared to the fault-free counterpart.

## I. INTRODUCTION

Graph Neural Networks (GNNs) have achieved state-of-the-art performance across a wide spectrum of graph-based applications such as node classification, link prediction, and graph clustering [1]. As a result, there is a growing demand for training GNNs at the edge. This necessitates the design of edge platforms based on single chips or embedded systems to support GNN training [2]. However, GNN training is both memory and compute-intensive. Conventional edge platforms, designed with CPUs/GPUs require large volumes of off-chip data movement, giving rise to performance bottlenecks [3]. Hence, this has motivated the need to explore computing architectures that reduce data movement for edge platforms. In this regard, processing-in-memory (PIM)-enabled architectures have emerged as a potential solution as they enable reduction of unnecessary data movement.

ReRAM-enabled PIM architectures have become popular for high-performance and energy-efficient Neural Network (NN) training and inferencing using edge devices with small form factors [4]. The crossbar array structure of ReRAM-based PIM architectures makes them well-suited for performing highly parallel matrix-vector multiplication (MVM) operations, which is the predominant computation kernel in both GNN training and inferencing. However, the relatively less mature fabrication process of ReRAMs compared to standard CMOS and their low endurance give rise to different types of hardware faults. These faults lead to unreliable training and poor test accuracy [5]. The most severe faults are stuck-at-faults (SAFs), which make the ReRAM cell resistance unchangeable. Hence, hardware/software fault-mitigation techniques are being studied to enable reliable training and testing using unreliable ReRAM-based systems.

The GNN training and inferencing computations are split into two phases: *aggregation* and *combination*. The aggregation phase computes the aggregated node features using the graph adjacency matrix, while the combination phase computes the node embeddings for the GNN layer using learnable weights. These two computation phases require that both graph adjacency matrices and GNN weights to be stored on ReRAM crossbars. However, existing fault-mitigation methods are mostly tailored towards NN with only their weight parameters mapped to ReRAM crossbars (such as in CNNs). As we show later, SAFs in ReRAM crossbars storing graph adjacency matrix also lead to significant accuracy loss. Hence, any fault-tolerant scheme should address faults on ReRAM cells storing both adjacency and weight matrices to ensure reliable GNN training without introducing significant performance overhead.

In this work, we propose a <u>Fa</u>ult-aware GNN training framework for <u>Re</u>RAM-based PIM architectures referred to as **FARe**. Our approach mitigates the adverse effect of faults in ReRAM crossbars when training with various GNN models and datasets. FARe considers the SAF distribution in ReRAM crossbars to appropriately map the graph adjacency matrix and leverages weight clipping to address faults on the GNN weight matrix. Unlike existing fault-tolerant approaches, the model- and dataset-agnostic nature of FARe makes it generalizable across different types of GNN workloads and graph datasets, which is demonstrated by our experiments. Specific contributions of this paper are:

- We propose a novel fault-tolerant framework called FARe that enables on-device GNN training using ReRAM-based architectures. FARe is agnostic to both GNN models and graph datasets.
- We demonstrate the limitations of existing fault-tolerant methods when used for GNN training, highlighting their significant performance overhead.
- We show that FARe achieves near-ideal accuracies in scenarios of high fault rates of up to 5%. Remarkably, FARe enables reliable training with <1% test accuracy loss and around 1% performance overhead with state-of-the-art GNN models and diverse graph datasets.

*To the best of our knowledge, this is the first work that comprehensively addresses SAFs in ReRAM-based PIM architectures for GNN training.* The rest of this paper is organized as follows; Section II discusses prior work; Section III outlines the GNN computation kernel, and Section IV elaborates on the proposed fault-tolerant solution. Section V presents the experimental results, and section VI concludes the paper by summarizing the key findings of this work.

## II. RELATED WORK

### A. Faults in ReRAMs

ReRAMs are susceptible to various types of faults. These faults lead to deviations in the resistance of a ReRAM cell due to various factors including noise, process variations, temperature, IR drop, etc. [5]. Among these, SAF is one of the most severe faults that can hinder reliable computation on ReRAM-based architectures [5]. SAFs manifest in two forms: Stuck-at-0 (SA0) and Stuck-at-1 (SA1) faults, causing the ReRAM cell to be permanently stuck in either a low resistance state (LRS) or a high resistance state (HRS) [6]. These faults an arise pre-deployment (at t = 0−) because of manufacturing

This work was supported, in part by the US National Science Foundation (NSF) under grants CSR-2308530, and CSR-1955353

defects or post-deployment (emerged during their use, i.e., at t > 0) due to limited cell endurance. Built-in self-test (BIST) circuits can identify the type and location of SAFs [7], which can be utilized to develop fault-tolerant design methodologies.

*B. Existing Fault-tolerant Techniques*

Several fault-tolerant techniques have been proposed to handle SAFs in ReRAM-based architectures. These techniques can be broadly categorized into hardware and software-based approaches. Hardware techniques typically involve adding redundancy to the system, such as using redundant columns as a replacement for faulty ReRAM columns [8]. The use of a fault map to compensate the output is also proposed as another possible solution [9]. However, these approaches require additional hardware, which increases the overall energy and area cost required for on-device training. Software approaches, on the other hand, typically involve implementing algorithmic mechanisms to mitigate the effects of SAFs. Neuron reordering is a remapping approach that proposes permutating neurons to overlap with SAFs for fault tolerance [7]. However, the performance impact of repeatedly computing the remapping during the training process is significant. Other software methods include model retraining or fine-tuning to recover the accuracy loss due to SAFs. Unstructured weight pruning fixes faulty weight elements to constant values based on SAF information [10]. Another retraining approach is stochastic training on pre-trained models [11]. However, fault-tolerant retraining algorithms are specifically targeted toward inferencing, and the focus of this paper is on training the GNN model from scratch. Recently, weight clipping has been proposed as a low-cost solution to deal with faults in ReRAM [12]. However, weight clipping cannot be used solely as a fault-mitigation technique, as faults affecting the adjacency matrix remain unaddressed by clipping only GNN weights. Table I summarizes the existing approaches aimed at SAF mitigation along with their capabilities and limitations. None of the existing methods have all the necessary features. Hence, we explore a new fault-tolerant framework for GNN training to fill this gap. This framework is aimed at handling SAFs in both phases of GNN computation while minimizing performance overheads.

### III. GNN COMPUTATION IN THE PRESENCE OF FAULTS

In this section, we discuss the impact of SAFs during GNN training on the ReRAM-based PIM accelerator.

*A. SAFs During Training*

Recently, ReRAM-based PIM architectures have been proposed to accelerate the sparse and dense MVM operations in the *aggregation* and *combination* phases (defined above) of GNN training and inferencing [13]. These architectures employ mini-batch training where the input graph is first partitioned into smaller subgraphs, and the subgraphs are processed in batches. Subsequently, a pipelined training strategy is adopted where all the GNN layers are processed simultaneously [3]. Consequently, both phases of GNN computation are susceptible to SAFs.

The weights on ReRAM-based architectures are commonly represented using 16-bit fixed-point precision. The 16 bits are distributed across multiple cells with architectures often adopting a 2-bit representation per cell. Subsequently, partial outputs are accumulated using a shift-and-add operation to obtain the final output of the MVM operation. Due to this distributed mapping, faults at different positions

TABLE I. COMPARISON OF EXISTING FAULT-TOLERANT TECHNIQUES

| Ref. | Training | Performance Overhead | Combination/ Aggregation | Mitigate Post-deployment Faults |
|---|---|---|---|---|
| [8] | Y | HIGH | Y / Y | Y |
| [10] | N | LOW | Y / N | N |
| [11] | N | LOW | Y / Y | N |
| [9] | N | HIGH | Y / N | N |
| [12] | Y | LOW | Y / N | Y |
| [7] | Y | HIGH | Y / Y | Y |

a. "Y" represents suitability, while "N" denotes unsuitability for the technique.

within a single weight vector have varying impacts on the final output. Specifically, faults near the most significant bit (MSB) position have an exponential effect when compared to faults near the least significant bit (LSB) position. Fig. 1(a) illustrates the effect of a SA1 fault near the MSB of the weight matrix. Unlike GNN weights, the adjacency matrix is stored in a binary format on crossbars indicating whether any given pair of vertices are connected in the graph or not. A SAF in the adjacency matrix can result in the addition (SA1) or deletion (SA0) of an edge in the stored graph, which alters the graph structure. SA0 faults in positions representing "one" in the adjacency matrix lead to edge deletion, whereas SA1 faults in positions representing "zero" adds an erroneous edge. The addition/deletion of edges results in incorrect *aggregation* leading to accuracy loss. Fig. 1(b) shows the effect of faults on the adjacency matrix.

### IV. THE FARe FRAMEWORK

In this section, we present the salient features of the proposed FARe framework. FARe uses a fault-tolerant mapping algorithm to map the graph adjacency matrix to faulty crossbars for the *aggregation* phase, and a weight clipping technique to mitigate the effects of ReRAM faults on the weights in the *combination* phase. These two synergistic strategies enable reliable GNN training on ReRAM-based architectures with both pre- and post-deployment faults.

*A. Handling Faults in The Aggregation Phase*

The *aggregation* phase in GNN training and inferencing performs neighborhood aggregation (or message passing) using the graph adjacency matrix. However, current techniques only focus on ensuring efficient hardware utilization while mapping the adjacency matrix and are oblivious to ReRAM faults [14]. In this work, we propose a fault-aware approach to map the adjacency matrix to ReRAM crossbars. Our approach guarantees efficient resource utilization while also incorporating fault-awareness. The algorithm reduces the problem to one of finding an optimal weighted bipartite graph matching that maps blocks of the adjacency matrix to ReRAM crossbars in a manner that maximizes its overlap with the location of SAFs, thereby minimizing the impact of faults. The static nature of the adjacency matrix facilitates a one-time computation to determine the fault-tolerant mapping considering pre-deployment faults.

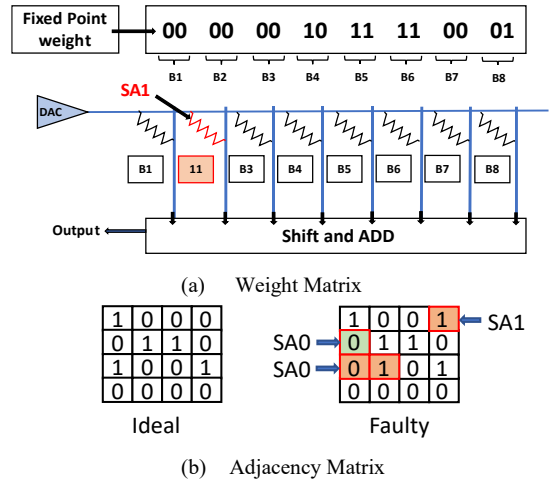

(a)  Weight Matrix

(b)  Adjacency Matrix

Fig.1. Conceptual illustration of SAF in the ReRAM crossbars storing the weigth and graph adjacency matrices.

The inputs to the mapping algorithm are: the adjacency matrix $A_i$ of the $i^{th}$ batch of subgraphs, the available set of $m$ crossbars ($C = \{c_0, c_1, ..., c_{m-1}\}$), and the existing ReRAM crossbar fault distribution ($\mathcal{F}[0, ..., m-1]$). The fault distribution $\mathcal{F}[j]$ of each crossbar $j$ is provided through a BIST circuit designed for SAF detection in ReRAM crossbars [7]. The BIST circuit adds a minimal area overhead (0.13% of total area) and is enabled initially to provide the pre-deployment fault map. The output of the algorithm is the fault-aware mapping ($\Pi$). Algorithm 1 shows the pseudocode for mapping the adjacency matrix $A_i$. The algorithm is executed during the pre-processing phase on the host device.

The steps of the mapping algorithm are as follows. First, the ($N \times N$) adjacency matrix $A_i$ is decomposed into a set $B$ of ($n \times n$) disjoint equal sized blocks, where $n$ is the number of rows (and columns) of an ReRAM crossbar ($n < N$). Next, we compute a $cost(i,j)$ function to map all the $n$ rows of block $a_i$ in $B$ onto any given crossbar $c_j$ in $C$ using its fault map $\mathcal{F}[j]$. This cost is defined as the minimum number of mismatches between a given row permutation of the matrix block and the target crossbar's SA0 and SA1 locations – e.g., the mapping example shown in Fig. 1(b) will generate a cost of 3. We formulate this mapping problem as a *weighted bipartite matching problem* (line 5 of Algorithm 1), by constructing a bipartite graph: $G_1(V_1, V_2, E)$ with $V_1$ as the set of $n$ rows of block $a_i$; $V_2$ as the set of $n$ rows of the crossbar $c_j$, and $E$ as the set of all edges connecting every row in $a_i$ to every row in $c_j$. The weight of an edge is the number of mismatched locations for that row-to-row mapping. In our implementation, we use the b-Suitor algorithm, which is a half-approximation algorithm for optimal matching [15].

Next, using the information of all possible $cost(i,j)$ values – i.e., minimum cost way to map every block in $B$ to every crossbar in $C$ – we determine an optimal assignment for mapping the $b$ blocks of $B$ to the $m$ crossbars of $C$, where $b \leq m$ (line 18 of Algorithm 1). Specifically, we consider another instance of minimum bipartite matching. The bipartite graph here is $G_2(V_1, V_2, E)$ with $V_1 = B, V_2 = C$, and $E$ comprising of all edges connecting blocks to crossbars with their corresponding $cost(i,j)$ weights attached. The minimum matching output so obtained ($\Pi$) is then an optimized way to map the blocks in $B$ given $m$ crossbars from the set $C$.

In addition, SA1 faults are more critical than SA0 faults as we show later. We incorporate this knowledge into the algorithm. In particular, we examine the minimum SA1 non-overlap resulting from mapping any block in $B$ to the crossbar $c_i$. Additionally, we compute the edge density of blocks in $B$ where edge density refers to the fraction of non-zero values (ones) within a given block. If the minimum non-overlap count is greater than the edge density of the sparsest block in $A_i$, we remove $c_i$ from the set of available crossbars (Line 12 of Algorithm 1). This reduces the value of $m$. In cases where $b = m$, we compute the density distribution of the adjacency matrix blocks in the batch. We observe edge density as low as

---

**Algorithm 1:** Algorithm for mapping an adjacency matrix on to the ReRAM crossbar

**Input:** $A_i[N, N]$: The adjacency matrix for batch $i$
$C = \{c_0, c_1, ... c_{m-1}\}$: The set of $m$ crossbars available for mapping
$\mathcal{F}[0, m-1]$: Fault distribution for all crossbars
$n$: The target number of rows per crossbar

**Output:** A mapping $\Pi$ of the blocks of $A_i$ on to $C$

1: $B \leftarrow$ Block decompose $A_i$ into blocks of size $n \times n$ each
2: Let $b \leftarrow |B|$
3: $cost[b, m] \leftarrow$ Initialize a 2D matrix
4: **for all** $\langle a_i, c_j \rangle \in \langle B, C \rangle$ **do**
5: $\quad cost[i,j] \leftarrow$ MinBipartiteMatch($a_i, c_j, \mathcal{F}_j$)
6: **end for**
7: $D_i \leftarrow$ the edge density of block $a_i, \forall a_i \in B$
8: **for** $j \in cost[:, m]$ **do**
9: $\quad a_{sp} \leftarrow D_i$.popMin() /* find sparsest block */
10: $\quad$ **if** min(SA1(cost[:,j])) > $a_{sp}$ **then**
11: $\quad\quad$ **if** $m > b$ **then**
12: $\quad\quad\quad$ remove $c_j$ from $C$
13: $\quad\quad$ **else**
14: $\quad\quad\quad$ remove $a_{sp}$ from $B$
15: $\quad\quad$ **end if**
16: $\quad$ **end if**
17: **end for**
18: $\Pi \leftarrow$ MinBipartiteMatch($B, C, cost$)
19: **return** $\Pi$

---

0.001 implying that only 0.1% of the values in the blocks are equal to one. Thus, we remove the sparsest block to be mapped onto crossbars (Line 14 of Algorithm 1) given the minimum non-overlap comparison. In such cases, we are reducing $b$ in our optimization problem to provide greater freedom while mapping. Note this is a worst-case scenario where SA1 fault pattern within a crossbar does not overlap with any blocks available in the batch even with row permutation.

**Post-deployment faults**: During the mini-batch GNN training, subgraph adjacency matrices need to be mapped to the ReRAM crossbars in batches to enable pipelined training [13] (as shown in Fig. 2). This process leads to multiple ReRAM cell write operations. Therefore, we also consider post-deployment faults in addition to pre-deployment faults during GNN training. To achieve this, we enable the BIST circuit at the end of each epoch to obtain the fault distribution due to post-deployment faults. The BIST circuit introduces a negligible timing overhead (~0.13% to the overall execution time). At the end of an epoch, Algorithm 1 can be reinitialized to address the post-deployment faults. However, the overall mapping will be similar, given endurance of ReRAM cells ($10^6$-$10^{12}$) is orders of magnitude higher than the number of writes in one epoch [5]. Hence, we only perform row permutation within crossbars to tackle post-deployment faults on top of the mapping $\Pi$ (line 18 of Algorithm 1) for few faults that might appear after an epoch. The necessary computation for the row permutations (bipartite matching) is performed on the host (CPU/GPU) where the bipartite graphs are already created while computing the parameter $\Pi$. The matching algorithm used for row-permutations has a linear time complexity and generates the mapping for the next batch parallelly on the host device while the current batch is being executed on ReRAMs. This ensures that mapping computation considering post-deployment faults does not lead to additional performance impact.

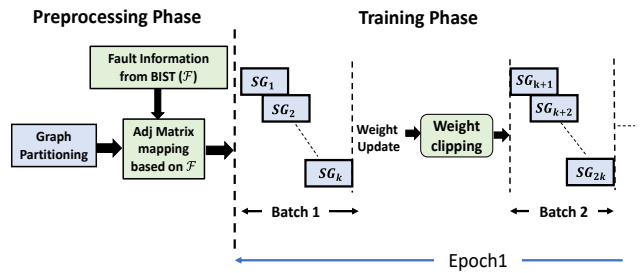

Fig.2. Pipelined GNN training in the presence of faults.

### B. Handling Faults in the Combination Phase

The *Combination* computation phase involves MVM with the GNN weight matrix. As discussed in Section III, a single SA1 fault near the MSB position can cause the weight value to become very high, also known as '*weight explosion*'. The explosion of even a small fraction of weights due to faults during training can cause significant deterioration of the predictive accuracy [12]. Therefore, it is crucial to tackle faults affecting weights, thereby ensuring reliable GNN training. We incorporate clipping during the *combination* phase to prevent weight explosion. The clipping operation restricts weight values from exceeding a specific threshold in presence of SAFs [12]. This threshold is a hyperparameter and remains constant throughout the training process. Applying clipping to unrealistically large weight values prevents the training process from becoming unstable. Consequently, the backpropagation algorithm can effectively train the remaining weights and compensate for the ones mapped to the faulty cells during training. This method acts as an implicit regularization technique and reduces the sensitivity of the loss function to faults. *Weight clipping is equally effective for both pre- and post-deployment faults without any modification.*

## V. EXPERIMENTAL RESULTS

In this section, we present an experimental analysis of the proposed FARe framework for fault-aware GNN training on ReRAM-based PIM architectures. First, we describe the experimental setup used for the performance evaluation. Next, we assess the impact of SA0/SA1 faults. Finally, we compare the effectiveness of FARe in terms of GNN model accuracy and overall performance with respect to existing methods.

### A. Experimental Setup

We consider four graph datasets in our experimental evaluation: protein-protein interaction (PPI), Reddit, Open Graph Benchmark-citation2 (Ogbl), and Amazon2M [16]. These graph datasets represent the typical size and diversity of use cases for training GNN models at the edge. Note that massive graphs (e.g., billions of nodes and edges) are best processed on cloud platforms and are not suitable for our work. We also use three types of GNN models to demonstrate the generalizability of FARe: Graph Convolution Networks (GCN), Graph Attention Networks (GAT), and Graph Sample and Aggregate (SAGE) [16]. Table II presents the details of the datasets, training hyperparameter configurations including learning rate (lr), batch size (Batch), number of graph partitions, number of training epochs used in this work. We employ mini-batch training, which uses small graph clusters (subgraphs) obtained from a monolithic graph via partitioning for GNN training [16]. Here, the graph is partitioned into smaller subgraphs using the METIS graph partitioning tool [17]. The partitioning step using METIS is a one-time process that takes a small portion of preprocessing time on the host device for the datasets considered here. The specifications of the ReRAM architecture used are described in Table III. Each ReRAM tile consumes 0.34W of power and occupies 0.157 mm$^2$ of area. The ReRAM tile is equipped with a 16-bit comparator and a 2:1 mux for implementing weight clipping, which has little area and power overheads [12]. The modeling of area, latency of all on-chip buffers, and peripheral circuits were obtained using NeuroSim v2.1 [18].

TABLE II. GRAPH DATASETS & GNN WORKLOAD CONFIGURATION

| Dataset | Dataset Statistics | | Hyperparameters (lr =0.01,epochs=100) | GNN Model |
|---|---|---|---|---|
| | # Nodes | # Edges | | |
| PPI | 56,944 | 818,716 | Batch=5, Partitions=250 | GCN |
| | | | | GAT |
| Reddit | 232,965 | 11,606,919 | Batch=10, Partitions=1500 | GCN |
| Amazon2M | 2,449,029 | 61,859,140 | Batch=20, Partitions=10,000 | GCN |
| | | | | SAGE |
| Ogbl | 2,927,963 | 30,561,187 | Batch=16, Partitions=15,000 | SAGE |

TABLE III. ReRAM-PIM ARCHITECTURE SPECIFICATIONS

| ReRAM Tile | 96-ADCs (8-bits), 12×128×8 DACs (1-bit), 96 crossbars, 128×128 crossbar size, 10MHz, 2-bit/cell resolution, 8-comparators (16-bit, 2GHz), 8-mux (2:1) |
|---|---|

We employ a PyTorch-based wrapper, which we incorporate into NeuroSim to simulate the effect of SAFs. SAFs are generally known to cluster across various fault centers [6]. Hence, we adopt a Poisson distribution of SAFs across the ReRAM crossbars and a uniform fault distribution within each crossbar. This results in some crossbars having a higher fault density than others, and an equal probability of fault occurrence in any ReRAM cell within a crossbar. Here, we define "fault density" as the percentage of ReRAM cells that are faulty in the architecture under consideration. Following prior work, we consider a SA0:SA1 fault density ratio of 9:1, indicating that SA0 faults are nine times more likely to occur than SA1 faults [6]. However, given the ongoing evolution of the ReRAM manufacturing process, we also consider a scenario where SA0 and SA1 faults have an equal probability of occurrence (i.e., SA0:SA1 = 1:1). FARe is equally applicable for any other fault ratio as well.

### B. Impact of SA0 and SA1 faults on GNN Training

First, we investigate the overall impact of SAFs on the GNN training process. Specifically, we assess the severity of SA0 and SA1 faults using two cases: i) SA0 only, and ii) SA1 only. We did not consider fault densities beyond 5% in our evaluation, as such densities are not practical in real-world scenarios. This is because any faulty chip with a fault density exceeding 5% can be tested offline and subsequently discarded. In Fig. 3, we show the impact of SA0/SA1 faults on the final test accuracy of a GNN trained with the SAGE model and Amazon2M dataset as an example. Faults were introduced into the crossbars storing the weights and adjacency matrices separately to assess the impact of faults on each computational phase. We observe that SA1 faults have a greater negative effect on accuracy than SA0 faults for both weight and adjacency matrices. This knowledge is integrated into FARe, which efficiently addresses SA1 faults during both phases (aggregation and combination) of GNN training.

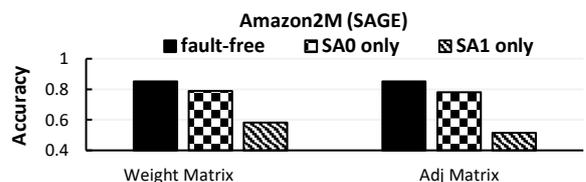

Fig.3. The accuracy of trained model after introducing 5% SA0 or SA1 pre-deployment faults on the weight and adjacency matrix separately.

*C. Training with faults*

Next, we evaluate the effectiveness of the proposed FARe framework during GNN training on faulty ReRAM crossbars. Specifically, we compare the training accuracy achieved by FARe (Fig. 4(b)) versus fault-unaware (Fig. 4(a)) on the Reddit dataset with the GCN model as an example. We observed the same trend for the remaining datasets and GNN models. Here, "fault-unaware" refers to the naïve implementation of GNN training *without* the incorporation of any fault-mitigation strategy. Additionally, for the baseline, we compared both FARe and fault-unaware against the fault-free GNN trained on ideal ReRAM crossbars. For a thorough evaluation, we vary the fault density from 1% to up to 5% with SA0:SA1 fault ratio of 9:1. Fig. 4(a) illustrates how the presence of faults seriously affects the accuracy for fault-unaware, making the training process unstable. Conversely, the training accuracy of the fault-aware GNN training using FARe overlaps with the fault-free training as the GNN model converges. This demonstrates the effectiveness of FARe in mitigating the adverse impact of faults in ReRAM crossbars.

*D. FARe Accuracy Evaluation*

Now, we evaluate the test accuracy of the trained models on faulty ReRAM crossbars using the FARe framework. We consider neuron reordering (NR) as one of the baselines since it can be used for both phases of GNN training unlike other techniques and does not need additional hardware (Table I). Since the neurons for the entire neural network are interconnected in a cascaded manner, we employ NR to permutate neurons to overlap with SAFs in both *combination* and *aggregation* phases in a unified manner. We also consider weight clipping as the second baseline to examine its effectiveness as a stand-alone fault-mitigation strategy during GNN training. Fig. 5 presents a comparative analysis of the test accuracy achieved by the trained GNN with three different models and four datasets when employing the fault-unaware method, NR, weight clipping, and FARe. In Fig. 5(a), we consider SA0:SA1 fault ratio of 9:1, and experiment with three fault densities: 1%, 3%, and 5%. As shown in Fig. 5(a), the use of fault-unaware mapping results in significant accuracy loss on faulty ReRAM crossbars, thereby demonstrating the need for a fault-tolerant approach. NR as a fault-tolerant technique improves accuracy; however, the accuracy drop is significant compared to the fault-free case. This is primarily due to the dimension of reordering being relatively large. For instance, considering a hidden layer with a dimension of 1024 and each GNN weight distributed across eight cells (using 2-bits/cell resolution, and 16-bit fixed point representation), the resulting reordering unit will have a dimension of 1024×8. Consequently, the use of such large units for reordering may not find significant overlap with SAFs. Similarly, weight clipping observes a significant accuracy drop as a stand-alone technique as faults affecting the aggregation phase of computation remain unaddressed. The faults in the adjacency matrix propagate false information leading to contamination in the learned GNN parameters after training. On the other hand, the FARe approach restores accuracy of the trained GNN with an accuracy drop of less than 1% even at high fault densities. This result demonstrates the remarkable effectiveness of FARe for fault-tolerant GNN training compared to the existing techniques.

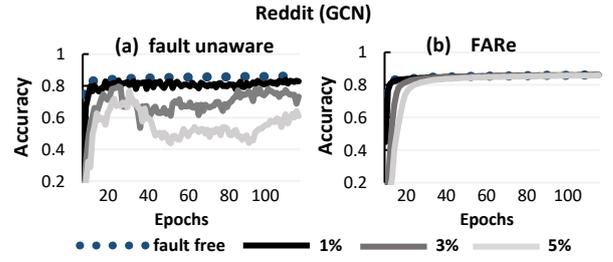

Fig.4. Training accuracy without/with the adoption of FARe approach under varying pre-deployment fault densities.

Now, we consider a scenario where SA1 and SA0 faults are equally probable. In Fig. 5(b), we show a similar accuracy comparison as earlier but with SA0:SA1=1:1. We observe a significantly higher accuracy drop with this fault ratio as an increase in the number of SA1 faults compared to the 9:1 SA0:SA1 ratio leads to a more substantial impact on the model's accuracy. Since FARe effectively addresses SA1 faults for both weights and adjacency matrices, it exhibits only a 1.1% loss in accuracy compared to fault-free case and restores the accuracy by 47.6% for the Reddit dataset as an example. Interestingly, NR performed especially poorly under 1:1 fault ratio, with an accuracy drop of 14.5%. The significant accuracy drop observed can be attributed to the fact that NR does not consider the criticality of SA1 fault over SA0 fault.

So far, we have considered only pre-deployment faults. Next, we also consider post-deployment faults during training.

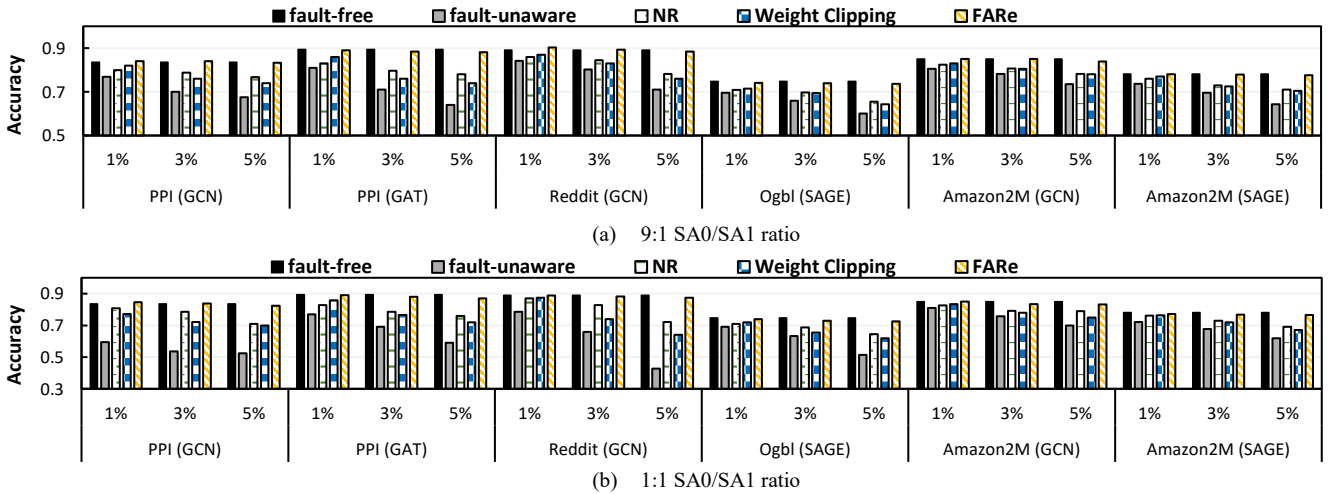

Fig.5. Comparative analysis of the trained model considering different pre-deployment fault densities, using the fault-unaware, NR, weight clipping, FARe approaches against the fault-free trained GNN model.

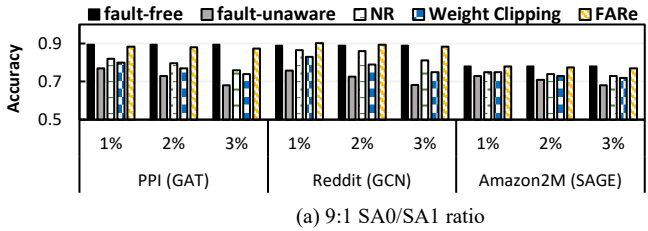
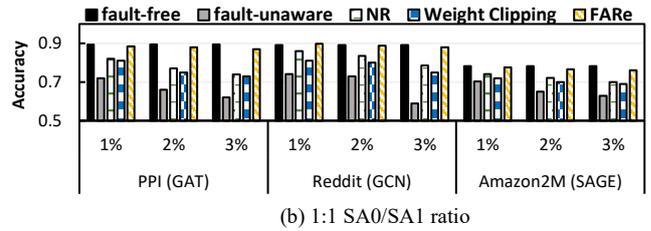

(a) 9:1 SA0/SA1 ratio  (b) 1:1 SA0/SA1 ratio

Fig.6. Comparative analysis of the trained model considering different pre-deployment fault densities along with 1% additional post-deployment faults, using the fault-unaware, NR, weight clipping, and FARe approaches against the fault-free trained GNN model.

To simulate post-deployment faults, we uniformly increase the fault density every epoch by a total of 1%, adding to the pre-deployment faults. *It is important to note that this represents a worst-case scenario, as faults may not occur after every epoch*. Fig. 6 shows the training accuracy considering post-deployment faults. For this experiment, we have considered three scenarios: 1%, 2%, and 3% pre-deployment fault density with 1% additional post-deployment faults. We experiment with SA0:SA1 fault ratios of both 9:1 and 1:1. We observe a similar trend where FARe demonstrates better fault tolerance with a maximum accuracy loss of 1.9%, while NR incurs an accuracy loss of up to 15%. Overall, FARe achieves nearly ideal accuracy levels, with an accuracy drop of less than 1% and 2% for SA0:SA1 fault ratios of 9:1 and 1:1 respectively across graph datasets and GNN models considering both pre- and post-deployment faults.

*E. FARe performance evaluation*

Next, we compare the timing overhead of using the fault-tolerant approaches with respect to training on fault-free hardware. The overall execution time of the pipelined implementation is determined by the *end-to-end pipeline depth* ($N + S - 1$) and the *delay of each pipeline stage*, where $N$ is the number of input subgraphs, and $S$ is the number of pipeline stages. Fig. 7 shows the normalized execution time for FARe, NR, and weight clipping. FARe introduces a 1% timing overhead initially in the pre-processing phase to determine the optimal mapping of the adjacency matrices on the crossbars. Further, weight clipping necessitates an additional pipeline stage to perform the clipping operation. Given that $N$ is significantly larger than $S$, the added cycle introduced by weight clipping is negligible in comparison to the total execution time. Overall, the execution of FARe introduces an overhead of ~1%. On the contrary, using NR as a fault-tolerant technique incurs a significant performance penalty. This is due to the pipeline being stalled repeatedly after processing each batch to perform reordering on the updated weight values.

## VI. CONCLUSION

ReRAM-based architectures enable high-performance and energy-efficient training of GNNs. However, the current ReRAM fabrication process and repeated writes result in faults that hinder reliable operation of the whole system. In this work, we have proposed a framework, FARe, to address faults in both the computational phases of GNN training.

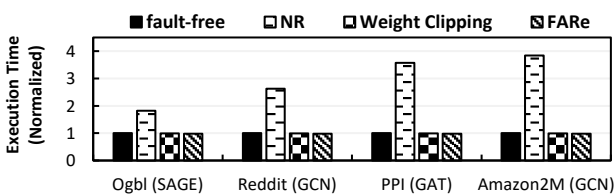

Fig.7. Normalized execution-time using FARe, NR, weight clipping, and w.r.t. fault-free training as the baseline.

FARe uses fault-aware mapping for the adjacency matrix along with weight clipping for the weight matrix to handle faults. Notably, FARe is agnostic to GNN models and achieves near-ideal accuracy even with a high fault density of 5%. Further, FARe achieves up to 4× speedup as compared to current fault-tolerant approaches.